\documentclass[12pt,english]{article}
\usepackage[T1]{fontenc}
\usepackage[latin1]{inputenc}
\usepackage{geometry}
\usepackage{amssymb,amsmath}

\usepackage{babel}
\usepackage{graphics}

\renewcommand{\thefootnote}{\fnsymbol{footnote}}
\newlength{\pubnumber} 

\catcode`\@=11
\@addtoreset{equation}{section}

\def\section{\@startsection{section}{1}{\z@}{3.5ex plus 1ex minus .2ex}
 {2.3ex plus .2ex}{\large\bf}}
\def\subsection{\@startsection{subsection}{2}{\z@}{2.3ex plus .2ex}
 {2.3ex plus .2ex}{\bf}}

\usepackage{amssymb}
\usepackage{epsfig}
\usepackage{cite}
\newcommand{\ba}{\begin{eqnarray}}
\newcommand{\ea}{\end{eqnarray}}

\begin{document}
\begin{titlepage}
\samepage{
\setcounter{page}{1}
\rightline{LTH--837}
\rightline{\tt arXiv:????.????}
\vfill
\begin{center}
 {\Large \bf  Little Heterotic Strings\\}
\vfill
 {\large  Alon E. Faraggi{}\footnote{
	E-mail address: faraggi@amtp.liv.ac.uk}
and 
	  Elisa Manno{}\footnote{
        E-mail address: manno@liverpool.ac.uk} 
\\
}
\vspace{.12in}
{} {\it           Department of Mathematical Sciences,
		University of Liverpool,     \\
                Liverpool L69 7ZL, United Kingdom}\\
\end{center}
\vfill
\begin{abstract}
  {\rm 
We discuss toroidal orbifolds of the $E_8\times E_8$ heterotic string,
in which the free fermionic Higgs--matter splitting is implemented
by a shift in the internal lattice coupled with the fermion numbers
of the gauge degrees of freedom. We consider models in which 
some choices of the orbifold result in the projection of the graviton. 
In the models that we consider the projection also results in flipping
the spin--statistics assignments in the massive string spectrum, 
whereas the massless spectrum
retains the conventional spin--statistics assignments. We argue 
that the partition functions are mathematically consistent for
one-- and multi--loop amplitudes, owning to the existence of supersymmetry
in the spectrum. 
A duality between different models at nonzero temperature is briefly discussed.
}
\end{abstract}
\smallskip}
\end{titlepage}

\renewcommand{\thefootnote}{\arabic{footnote}}
\setcounter{footnote}{0}

\def\l{\label}
\def\beq{\begin{equation}}
\def\eeq{\end{equation}}
\def\beqn{\begin{eqnarray}}
\def\eeqn{\end{eqnarray}}

\def\ie{{\it i.e.}}
\def\eg{{\it e.g.}}
\def\half{{\textstyle{1\over 2}}}
\def\third{{\textstyle {1\over3}}}
\def\quarter{{\textstyle {1\over4}}}
\def\m{{\tt -}}
\def\p{{\tt +}}

\def\slash#1{#1\hskip-6pt/\hskip6pt}
\def\slk{\slash{k}}
\def\GeV{\,{\rm GeV}}
\def\TeV{\,{\rm TeV}}
\def\y{\,{\rm y}}
\def\SM{Standard-Model }
\def\SUSY{supersymmetry }
\def\SSSM{supersymmetric standard model}
\def\vev#1{\left\langle #1\right\rangle}
\def\l{\langle}
\def\r{\rangle}

\def\Htw{{\tilde H}}
\def\chibar{{\overline{\chi}}}
\def\qbar{{\overline{q}}}
\def\ibar{{\overline{\imath}}}
\def\jbar{{\overline{\jmath}}}
\def\Hbar{{\overline{H}}}
\def\Qbar{{\overline{Q}}}
\def\abar{{\overline{a}}}
\def\alphabar{{\overline{\alpha}}}
\def\betabar{{\overline{\beta}}}
\def\tautwo{{ \tau_2 }}
\def\thetatwo{{ \vartheta_2 }}
\def\thetathree{{ \vartheta_3 }}
\def\thetafour{{ \vartheta_4 }}
\def\ttwo{{\vartheta_2}}
\def\tthree{{\vartheta_3}}
\def\tfour{{\vartheta_4}}
\def\ti{{\vartheta_i}}
\def\tj{{\vartheta_j}}
\def\tk{{\vartheta_k}}
\def\calF{{\cal F}}
\def\smallmatrix#1#2#3#4{{ {{#1}~{#2}\choose{#3}~{#4}} }}
\def\ab{{\alpha\beta}}
\def\Minv{{ (M^{-1}_\ab)_{ij} }}
\def\bone{{\bf 1}}
\def\ii{{(i)}}
\def\V{{\bf V}}
\def\b{{\bf b}}
\def\N{{\bf N}}
\def\t#1#2{{ \Theta\left\lbrack \matrix{ {#1}\cr {#2}\cr }\right\rbrack }}
\def\C#1#2{{ C\left\lbrack \matrix{ {#1}\cr {#2}\cr }\right\rbrack }}
\def\tp#1#2{{ \Theta'\left\lbrack \matrix{ {#1}\cr {#2}\cr }\right\rbrack }}
\def\tpp#1#2{{ \Theta''\left\lbrack \matrix{ {#1}\cr {#2}\cr }\right\rbrack }}
\def\l{\langle}
\def\r{\rangle}

\def\La{\Lambda}
\def\te{\theta}


\def\inbar{\,\vrule height1.5ex width.4pt depth0pt}

\def\IC{\relax\hbox{$\inbar\kern-.3em{\rm C}$}}
\def\IQ{\relax\hbox{$\inbar\kern-.3em{\rm Q}$}}
\def\IR{\relax{\rm I\kern-.18em R}}
 \font\cmss=cmss10 \font\cmsss=cmss10 at 7pt
\def\IZ{\relax\ifmmode\mathchoice
 {\hbox{\cmss Z\kern-.4em Z}}{\hbox{\cmss Z\kern-.4em Z}}
 {\lower.9pt\hbox{\cmsss Z\kern-.4em Z}}
 {\lower1.2pt\hbox{\cmsss Z\kern-.4em Z}}\else{\cmss Z\kern-.4em Z}\fi}

\def\AEF{A.E. Faraggi}
\def\NPB#1#2#3{{Nucl.\ Phys.}\/ {B \bf #1} (#2) #3}
\def\PLB#1#2#3{{Phys.\ Lett.}\/ {B \bf #1} (#2) #3}
\def\PRD#1#2#3{{Phys.\ Rev.}\/ {D \bf #1} (#2) #3}
\def\PRL#1#2#3{{Phys.\ Rev.\ Lett.}\/ {\bf #1} (#2) #3}
\def\PRP#1#2#3{{Phys.\ Rep.}\/ {\bf#1} (#2) #3}
\def\MODA#1#2#3{{Mod.\ Phys.\ Lett.}\/ {\bf A#1} (#2) #3}
\def\IJMP#1#2#3{{Int.\ J.\ Mod.\ Phys.}\/ {A \bf #1} (#2) #3}
\def\nuvc#1#2#3{{Nuovo Cimento}\/ {\bf #1A} (#2) #3}
\def\JHEP#1#2#3{{JHEP} {\textbf #1}, (#2) #3}
\def\EJP#1#2#3{{\it Eur.\ Phys.\ Jour.}\/ {\bf C#1} (#2) #3}
\def\MPLA#1#2#3{{\it Mod.\ Phys.\ Lett.}\/ {\bf A#1} (#2) #3}
\def\IJMPA#1#2#3{{\it Int.\ J.\ Mod.\ Phys.}\/ {\bf A#1} (#2) #3}

\def\etal{{\it et al\/}}

\hyphenation{su-per-sym-met-ric non-su-per-sym-met-ric}
\hyphenation{space-time-super-sym-met-ric}
\hyphenation{mod-u-lar mod-u-lar--in-var-i-ant}


\setcounter{footnote}{0}
\section{Introduction}
\bigskip

The Standard Model of particle physics is a non--supersymmetric 
renormalizable point--like quantum field theory and it accounts 
successfully for all sub--atomic experimental data to date.
On the other hand, gravity, which describes physical phenomena 
at the celestial and cosmological scales cannot be formulated
consistently as a non--supersymmetric point like quantum field 
theory. However, there is nothing sacred about the point--like 
idealisation of elementary particles.
The mundane generalisation of point--like particles to strings provides 
a framework for a consistent common formalism of gravity and 
the sub--atomic interactions. Indeed, string theory gives rise to
phenomenological models that are used to explore how string theory
connects to contemporary experimental data.
An important step in
the development of string theory
was obtained by the understanding that the five different
supersymmetric ten dimensional theories, as well as
eleven dimensional supergravity, may be related by perturbative
and nonperturbative duality transformations
\cite{Hull:1994ys,Witten:1995ex}.
However, it is well known that string theory
gives rise to a number of nonsupersymmetric ten dimensional vacua
that may be tachyonic or non--tachyonic 
\cite{Dixon:1986iz,lewellen}. 
It is plausible that the
resolution of some of the phenomenological issues facing string 
theory, in particular in relation to the cosmological evolution and
vacuum selection, will be gained by improved understanding 
of these non--supersymmetric vacua and how they fit in the 
fundamental theory that underlies the string theories. Furthermore,
it is likely that progress can be achieved by exploring some of the
features of these vacua in connection with the phenomenological string
vacua. 

In this paper we therefore pursue this line of exploration. The 
relevant class of phenomenological string vacua are 
the quasi--realistic heterotic--string 
models in the free fermionic formulation that are
related to $Z_2\times Z_2$ orbifold models. 
Quasi--realistic orbifold models have been constructed primarily by using
bosonic techniques (see {\it e.g.} \cite{nillesraby} and references therein), 
that provide a better insight to the underlying geometry. 
There is crucial distinction, however, between the models based on
the fermionic techniques and those based on the bosonic techniques 
in that the latter start from the $E_8\times E_8$ heterotic--string 
and compactify six dimensions on an internal manifold, whereas the 
fermionic models are formulated directly in four space--time dimensions.
In particular, the fermionic models utilise a projection that is reminiscent
of the projection from the ten dimensional 
$E_8\times E_8$ vacuum to the $SO(16)\times SO(16)$ one.
The important consequence of this projection in the 
quasi--realistic fermionic orbifold
models is that it  breaks the $E_6$ GUT symmetry to $SO(10)\times U(1)$
and splits the spinorial (matter) and vectorial (Higgs)
representations \cite{higgsmattersplit}. 
In ref. \cite{partitions} it was shown how the 
two partition functions pertaining to the four dimensional free fermionic
models with $E_8\times E_8$ and $SO(16)\times SO(16)$ gauge groups can
be connected by an orbifold projection. A route to improve the
understanding of phenomenological string models
can be achieved by implementing this projection in the corresponding
bosonic orbifold constructions. In this paper we undertake this task.
In the pursuit of this analysis we will explore a slight variation 
of our initial orbifold that will open up new vistas. The slight variation
is a discrete change of a sign in the orbifold action that results in 
a partition function that does not contain gravity at all, {\it i.e.} it 
is a ``little heterotic string'',
 in this sense the model under consideration  is similar
to the little string theories considered previously
(see for example 
\cite{Kutasov:2000jp,Aharony:1999ks,Kutasov:2001uf}
and references therein).
Yet, it corresponds to a modular invariant
critical string theory, {\it i.e.} it is a consistent string vacuum. 
In the spirit that we espouse here, in order to improve our understanding 
of string theory, it is vital to explore all the consistent string vacua
and not merely those that may be phenomenologically relevant. 

While the orbifold projection is a discrete choice, 
it is often seen that it can arise 
from discrete choices of vacuum expectation values of background fields. 
In this respect the a priory distinct vacua may be connected by continuous
variation of the background field. While in the effective field theory limit
the two vacua are distinct and cannot be connected, at the 
string level massless and massive states are 
exchanged by duality transformations. This is the view for 
example in the case of the spinor--vector duality recently observed
in free fermionic heterotic string models \cite{spduality}.
Following this lead and by compactifying the time coordinate we
investigate the possibility that the two partition functions can arise
from suitable choices of the background parameters and argue that this is 
in fact possible. The implication is that the graviton is not different
from any other string mode. Namely, it can be exchanged between the massive
and massless spectrum and can be viewed as arising dynamically by some
suitable choice of specific background parameters. 

\section{Elements of free fermionic models}

In this section we highlight the features of the free fermion models
that will be 
exploited in subsequent analysis. In the free fermionic formulation
of the heterotic
string in four dimensions all the world-sheet
degrees of freedom  required to cancel
the conformal anomaly are represented in terms of free fermions
propagating on the string world-sheet \cite{fff}.
In the light-cone gauge the world-sheet fermion field content consists
of twenty real fermions in the supersymmetric sector and forty--four real 
fermions in the non--supersymmetric sector.   
Under parallel transport around a non-contractible loop on the toroidal
world-sheet the fermionic fields pick up a phase,
$
f~\rightarrow~-{\rm e}^{i\pi\alpha(f)}f~,~~\alpha(f)\in(-1,+1].
$
Each set of specified
phases for all world-sheet fermions around all the non-contractible
loops is called the spin structure of the model. Such spin structures
are usually given is the form of 64 dimensional boundary condition vectors,
with each element of the vector specifying the phase of the corresponding
world-sheet fermion. The basis vectors are constrained by string consistency
requirements and completely determine the vacuum structure of the model.
The physical spectrum is obtained by applying the generalised GSO projections.

The boundary condition basis defining a typical 
``realistic free fermionic heterotic string model'' is 
constructed in two stages. 
The first stage consists of the NAHE set,
which is a set of five boundary condition basis vectors, 
$\{ 1 ,S,b_1,b_2,b_3\}$ \cite{nahe}. 
The gauge group after imposing the GSO projections induced
by the NAHE set is ${\rm SO} (10)\times {\rm SO}(6)^3\times {\rm E}_8$
with ${N}=1$ supersymmetry. The space-time vector bosons that generate
the gauge group arise from the Neveu-Schwarz sector and
from the sector $\xi_2\equiv 1+b_1+b_2+b_3$. The Neveu-Schwarz sector
produces the generators of ${\rm SO}(10)\times {\rm SO}(6)^3\times 
{\rm SO}(16)$. The $\xi_2$-sector produces the spinorial 128
of SO(16) and completes the hidden gauge group to ${\rm E}_8$.
The second stage of the
construction consists of adding to the 
NAHE set three (or four) additional boundary condition basis vectors,
typically denoted by $\{\alpha,\beta,\gamma\}$ \cite{ffmreview}. 
These additional basis vectors reduce the number of generations
to three chiral generations, one from each of the sectors $b_1$,
$b_2$ and $b_3$ and simultaneously break $SO(10)$ to one of its subgroups
${\rm SU}(5)\times {\rm U}(1)$, ${\rm SO}(6)\times {\rm SO}(4)$,
${\rm SU}(3)\times {\rm SU}(2)\times {\rm U}(1)^2$ 
or ${\rm SU}(3)\times {\rm SU}(2)^2\times {\rm U}(1)$.

The correspondence of the NAHE-based free fermionic models
with the orbifold construction is illustrated
by extending the NAHE set, $\{ 1,S,b_1,b_2,b_3\}$, by one additional
boundary condition basis vector \cite{foc}
\beq
\xi_1=(0,\cdots,0\vert{\underbrace{1,\cdots,1}_{{\bar\psi^{1,\cdots,5}},
{\bar\eta^{1,2,3}}}},0,\cdots,0)~.
\label{vectorx}
\eeq
With a suitable choice of the GSO projection coefficients the 
model possesses an ${\rm SO}(4)^3\times {\rm E}_6\times {\rm U}(1)^2
\times {\rm E}_8$ gauge group
and ${N}=1$ space-time supersymmetry. The matter fields
include 24 generations in the 27 representation of
${\rm E}_6$, eight from each of the sectors $b_1\oplus b_1+\xi_1$,
$b_2\oplus b_2+\xi_1$ and $b_3\oplus b_3+\xi_1$.
Three additional 27 and $\overline{27}$ pairs are obtained
from the Neveu-Schwarz $\oplus~\xi_1$ sector.

To construct the model in the orbifold formulation one starts
with the compactification on a torus with nontrivial background
fields \cite{Narain}.
The subset of basis vectors
\beq
\{ 1,S,\xi_1,\xi_2\}
\label{neq4set}
\eeq
generates a toroidally-compactified model with ${N}=4$ space-time
supersymmetry and ${\rm SO}(12)\times {\rm E}_8\times {\rm E}_8$ gauge group.
The same model is obtained in the geometric (bosonic) language
by tuning the background fields to the values corresponding to
the SO(12) lattice. The 
metric of the six-dimensional compactified
manifold is then the Cartan matrix of SO(12), 
while the antisymmetric tensor is given by
\beq
B_{ij}=
\begin{cases}
G_{ij}&;\ i>j,\cr
0&;\ i=j,\cr
-G_{ij}&;\ i<j.
\end{cases}
\label{bso12}
\eeq
When all the radii of the six-dimensional compactified
manifold are fixed at $R_I=\sqrt2$, it is seen that the
left- and right-moving momenta
$
P^I_{R,L}=[m_i-{1\over2}(B_{ij}{\pm}G_{ij})n_j]{e_i^I}^*
$
reproduce the massless root vectors in the lattice of
SO(12). Here $e^i=\{e_i^I\}$ are six linearly-independent
vielbeins normalised so that $(e_i)^2=2$.
The ${e_i^I}^*$ are dual to the $e_i$, with
$e_i^*\cdot e_j=\delta_{ij}$.

Adding the two basis vectors $b_1$ and $b_2$ to the set
(\ref{neq4set}) corresponds to the ${Z}_2\times {Z}_2$
orbifold model with standard embedding.
Starting from the Narain model with ${\rm SO}(12)\times 
{\rm E}_8\times {\rm E}_8$
symmetry~\cite{Narain} and applying the ${Z}_2\times {Z}_2$ 
twist on the
internal coordinates, reproduces
the spectrum of the free-fermion model
with the six-dimensional basis set
$\{ 1,S,\xi_1,\xi_2,b_1,b_2\}$.
The Euler characteristic of this model is 48 with $h_{11}=27$ and
$h_{21}=3$.
It is noted that the effect of the additional basis vector $\xi_1$ of 
eq. (\ref{vectorx}) is to separate the gauge degrees of freedom, spanned by
the world-sheet fermions $\{{\bar\psi}^{1,\cdots,5},
{\bar\eta}^{1},{\bar\eta}^{2},{\bar\eta}^{3},{\bar\phi}^{1,\cdots,8}\}$,
from the internal compactified degrees of freedom $\{y,\omega\vert
{\bar y},{\bar\omega}\}^{1,\cdots,6}$. 
In the "realistic free fermionic
models" this is achieved by the vector $2\gamma$ \cite{foc}
\beq
2\gamma=(0,\cdots,0\vert{\underbrace{1,\cdots,1}_{{\bar\psi^{1,\cdots,5}},
{\bar\eta^{1,2,3}} {\bar\phi}^{1,\cdots,4}} },0,\cdots,0)~,
\label{vector2gamma}
\eeq
which breaks the ${\rm E}_8\times {\rm E}_8$ symmetry to ${\rm SO}(16)\times 
{\rm SO}(16)$. 
The ${Z}_2\times {Z}_2$ twist breaks the gauge symmetry to
${\rm SO}(4)^3\times {\rm SO}(10)\times {\rm U}(1)^3\times {\rm SO}(16)$.
The orbifold still yields a model with 24 generations,
eight from each twisted sector,
but now the generations are in the chiral 16 representation
of SO(10), rather than in the 27 of ${\rm E}_6$. The same model can
be realized with the set
$\{ 1,S,\xi_1,\xi_2,b_1,b_2\}$,
by projecting out the $16\oplus{\overline{16}}$
from the $\xi_1$-sector by taking
\beq
c{\xi_1\choose \xi_2}\rightarrow -c{\xi_1\choose \xi_2}.
\label{changec}
\eeq
This choice also projects out the massless vector bosons in the
128 of SO(16) in the hidden-sector ${\rm E}_8$ gauge group, thereby
breaking the ${\rm E}_6\times {\rm E}_8$ symmetry to
${\rm SO}(10)\times {\rm U}(1)\times {\rm SO}(16)$.
The freedom in ({\ref{changec}) corresponds to 
a discrete torsion in the toroidal orbifold model.
At the level of the ${N}=4$ 
Narain model generated by the set (\ref{neq4set}),
we can define two models, ${Z}_+$ and ${Z}_-$, depending on the sign
of the discrete torsion in eq. (\ref{changec}). The first, say ${Z}_+$,
produces the ${\rm E}_8\times {\rm E}_8$ model, whereas the second, say 
${Z}_-$, produces the ${\rm SO}(16)\times {\rm SO}(16)$ model. 
However, the ${Z}_2\times 
{Z}_2$
twist acts identically in the two models and their physical characteristics
differ only due to the discrete torsion eq. (\ref{changec}). 
The partition functions corresponding to the $Z_{-}$ and $Z_{+}$
vacua are given respectively by 
\beqn
{Z}_- = ({V}_8-{S}_8) &\times&\left[~\left( |O_{12}|^2~+~|V_{12}|^2 ~\right) 
\left( \overline O_{16} \overline  O_{16}+  \overline C_{16}  
\overline C_{16}\right)\right.
\cr
&& + \left( |S_{12}|^2~~+|C_{12}|^2 ~\right) \left(  \overline S_{16} 
\overline S_{16}+
 \overline V_{16}  \overline V_{16}\right)
\cr
&& + \left( O_{12} \overline  V_{12} + V_{12} \overline  O_{12} \right)
\left(  \overline S_{16}  \overline V_{16} +  \overline V_{16}  
\overline S_{16}\right)
\cr
&& + \left. \left( S_{12}  \overline C_{12} +C_{12}  \overline S_{12} \right)
\left(  \overline O_{16}  \overline C_{16} +  
\overline C_{16}  \overline O_{16} \right) \right] \,
\label{zminus}
\eeqn
and
\beq
{Z}_+=( V_8- S_8)\left[|O_{12}|^2+|V_{12}|^2+|S_{12}|^2+|C_{12}|^2\right]
\left( \overline  O_{16} + \overline S_{16}\right) \left(  \overline O_{16} +  
\overline S_{16}\right) \,,
\label{zplus}
\eeq
depending on the sign of the discrete torsion in eq. (\ref{changec}). 
Here we have written ${Z}_{\pm}$ in terms of level-one
${\rm SO} (2n)$ characters (see, for instance \cite{Angelantonj:2002ct})
\beqn
O_{2n} &=& {\textstyle{1\over 2}} \left( {\vartheta_3^n \over \eta^n} +
{\vartheta_4^n \over \eta^n}\right) \,,
\nonumber \\
V_{2n} &=& {\textstyle{1\over 2}} \left( {\vartheta_3^n \over \eta^n} -
{\vartheta_4^n \over \eta^n}\right) \,,
\nonumber \\
S_{2n} &=& {\textstyle{1\over 2}} \left( {\vartheta_2^n \over \eta^n} +
i^{-n} {\vartheta_1^n \over \eta^n} \right) \,,
\nonumber \\
C_{2n} &=& {\textstyle{1\over 2}} \left( {\vartheta_2^n \over \eta^n} -
i^{-n} {\vartheta_1^n \over \eta^n} \right) \,.
\label{thetacharacters}
\eeqn

These two models can actually be connected by the orbifold
\beq
{Z}_- = {Z}_+ / a \otimes b \,,
\label{zminusfromzplus}
\eeq
with
\beqn
a &=& (-1)^{F_{\rm L}^{\rm int} + F_{\xi_1}} \,,
\nonumber \\
b &=& (-1)^{F_{\rm L}^{\rm int} + F_{\xi_2} }\,, \label{orbzpm}
\eeqn
where $F_L^{\rm int}$ is a fermion number in the internal lattice and 
$F_{\xi_1}$, $F_{\xi_2}$ are fermion numbers acting in the observable and
hidden gauge sectors, respectively.
The orbifold projection given in eqs (\ref{zminusfromzplus}) and (\ref{orbzpm})
is defined at the free fermionic point in the moduli space
since $Z_+$ and $Z_-$ are expressed at this point. However, it can be
generalised to arbitrary points in the moduli space and hence can be
employed to construct orbifold models that originate from the $Z_-$ 
partition function, in analogy to the case in the free fermionic
constructions. 
Let us consider for simplicity the case of six orthogonal circles with
radii $R_i$. The partition function reads
\beq
{ Z}_+ = ( V_8 - S_8) \, \left( \sum_{m,n} \Lambda_{m,n}
\right)^{\otimes 6}\, \left(  \overline O _{16} + \overline S_{16} \right) 
\left( \overline O _{16} +  \overline S_{16} \right)\,,
\label{e8xe8partin4d}
\eeq
where as usual, for each circle,
\beq
p_{\rm L,R}^i = {m_i \over R_i} \pm {n_i R_i \over \alpha '} \,
\eeq
and
\beq
\Lambda_{m,n} = {q^{{\alpha ' \over 4} 
p_{\rm L}^2} \, \bar q ^{{\alpha ' \over 4} p_{\rm R}^2} \over |\eta|^2}\,.
\eeq

In the case of one compactified dimension the $Z_+$ partition function is 
given by 
\beq
{ Z}_+^{9d} = ( V_8 -  S_8) \,  \Lambda_{m,n} \,
		\left( \overline  O_{16} +  \overline S_{16} \right)
		\left(  \overline O_{16} +  \overline S_{16} \right)\,.
\label{zplusin9d}
\eeq
Applying the orbifold projections 
\beqn
a &=& (-1)^{F_{\xi_1}}\delta \,, \nonumber\\
b &=& (-1)^{F_{\xi_2}}\delta \,,
\label{deltaorbifold}
\eeqn
where 
$\delta x^9 = x_9 +\pi R$,
${\xi_1}=\{{\bar\psi}^{1,\cdots,5}, {\bar\eta}^{1,2,3}\}$ 
and
${\xi_2}=\{{\bar\phi}^{1,\cdots,8}\}$, 
in $Z_+^{9d}$ produces the $Z_-^{9d}$ partition function given by
\beqn
{ Z}_-^{9d} = \sqrt{\tau_2}Z_{bos.} (V_8 - S_8)~\left[ \right. 
& & \left. \Lambda_{2m,n} \,~~~~~~\left(\overline O_{16}\overline O_{16}+
\overline C_{16}\overline C_{16}\right) \right. \nonumber\\
 &+& \left. \Lambda_{2m+1,n} \, ~~~\left(\overline S_{16}\overline S_{16}~+
\overline V_{16}\overline V_{16}\right)\right. \nonumber\\
 &+& \left. \Lambda_{2m,n+{1\over2}} \, ~~~\left(\overline S_{16}\overline 
V_{16}+\overline V_{16}\overline S_{16}\right)\right. \nonumber\\
 &+&  \left. \Lambda_{2m+1,n+{1\over2}} \, \left(\overline O_{16}\overline 
C_{16}+\overline C_{16}\overline O_{16}\right)\right]~.\nonumber\\
\label{zminus9d}
\eeqn
We note here that the shift given by $\delta$ differs from the shifts
that were 
found in ref. \cite{partitions} to reproduce the partition function of the
$SO(12)$ lattice at the maximally symmetric 
free fermionic point, given in eq.(\ref{zplus}). Since the string on a
compactified coordinate contains momentum and winding modes, one can
shift the coordinate along either and also allow shifts that mix the momentum
and winding modes. Indeed, the precise identification of the lattice at the 
free fermionic point is obtained for shifts that mix momentum and winding
modes. Here, we restrict ourselves to the simpler shifts and incorporation
of other shifts is left for future work.

The partition function from the free fermion model $\{1,S,\xi_1,\xi_2\}$, 
with $1+S+\xi_1+\xi_2=\{y^1 ,\omega^1~| ~\bar{y}^1,\bar{\omega}^1\}$, 
is given by\footnote{We included the terms with $\theta_1$ to facilitate
the translation to the characters given in eq. (\ref{thetacharacters}).}
\beqn 
Z_{9d}&=&{1\over 2^4}
\left( \theta_3^4-\theta_4^4-\theta_2^4-\theta_1^4\right) 
\left\{
\left(|\theta_3|^2+|\theta_4|^2+|\theta_2|^2+|\theta_1|^2\right)
\left(\bar{\theta}_3^{16}+\bar{\theta}_4^{16} + \bar{\theta}_2^{16} 
+ \bar{\theta}_1^{16}\right)\right.
\cr
\cr
&&+\left[|\theta_3|^2+|\theta_4|^2+c{\xi_1\choose \xi_2}
\left(|\theta_2|^2+|\theta_1|^2\right)\right]
\left[\bar{\theta}_3^8\bar{\theta}_4^8+
\bar{\theta}_4^8\bar{\theta}_3^8+c{\xi_1\choose \xi_2}
\left(\bar{\theta}_2^8\bar{\theta}_1^8+
\bar{\theta}_1^8\bar{\theta}_2^8\right)\right]
\cr
\cr
&&+\left[|\theta_3|^2+c{\xi_1\choose \xi_2}
\left(|\theta_4|^2+|\theta_1|^2\right)+|\theta_2|^2\right]
\left[\bar{\theta}_2^8\bar{\theta}_3^8+\bar{\theta}_3^8\bar{\theta}_2^8+
c{\xi_1\choose \xi_2}\left(\bar{\theta}_4^8\bar{\theta}_1^8+
\bar{\theta}_1^8\bar{\theta}_4^8\right)\right]
\cr
\cr
&&\left.\left[c{\xi_1\choose \xi_2}
\left(|\theta_3|^2+|\theta_1|^2\right)+\theta_4|^2+|\theta_3|^2\right]
\left[\bar{\theta}_2^8\bar{\theta}_4^8+\bar{\theta}_4^8\bar{\theta}_2^8+
c{\xi_1\choose \xi_2}\left(\bar{\theta}_3^8\bar{\theta}_1^8+
\bar{\theta}_1^8\bar{\theta}_3^8\right)\right]\right\}.\nonumber\\
\ea
In terms of the characters given in eq. (\ref{thetacharacters}) we have
\beq
 Z_+=(V_8-S_8)\left(|O_2|^2+|V_2|^2+|S_2|^2+|C_2|^2\right)
\left( \overline O_{16} + \overline S_{16}\right) \left( \overline O_{16} +
\overline S_{16}\right) 
\eeq
and
\ba
 Z_-= (V_8-S_8)&\times& \left[\left(|O_2|^2+|V_2|^2\right)
\left(\overline O_{16} \overline O_{16}+ \overline C_{16} 
\overline C_{16}\right)
\right. \cr\cr
&& + 
\left(|S_2|^2+|C_2|^2\right) \left( \overline S_{16} \overline S_{16}+
\overline V_{16} \overline V_{16}\right) 
\cr
\cr
&& + \left(O_2\overline{V}_2+V_2\overline{O}_2\right)
\left( \overline S_{16} \overline V_{16} + \overline V_{16} 
\overline S_{16}\right) \cr\cr
&&  + \left. \left(S_2\overline{C}_2+C_2\overline{S}_2\right)
\left( \overline O_{16} \overline C_{16} + 
\overline C_{16} \overline O_{16} \right) \right] .
\ea

In this case the orbifold operation is 
\beqn
 a & = & (-1)^{F_L^{\rm int}+F_{\xi_1}} \,, \nonumber\\ 
 b & = & (-1)^{F_L^{\rm int}+F_{\xi_2}} \,, \label{9dspecial}
\eeqn
where $F_L^{\rm int}$ acts on  $9^{\text{th}}$ dimension.

\subsection{10D case} 
We comment here on the ten dimensional case. 
The free fermionic model is $\{1,\xi_1,\xi_2\}$, with $1+\xi_1+\xi_2=S$.
\ba 
Z_{10d}&=&{1\over 2^3}\left\{\left( \theta_3^4-\theta_4^4-
\theta_2^4-\theta_1^4\right) \left(\bar{\theta}_3^{16}+\bar{\theta}_4^{16} 
+ \bar{\theta}_2^{16} + \bar{\theta}_1^{16}\right)\right.
\cr
\cr
&&+\left[\theta_3^4-\theta_4^4-c{\xi_1\choose \xi_2}
\left(\theta_2^4+\theta_1^4\right)\right]
\left[\bar{\theta}_3^8\bar{\theta}_4^8+
\bar{\theta}_4^8\bar{\theta}_3^8+
c{\xi_1\choose \xi_2}\left(\bar{\theta}_2^8\bar{\theta}_1^8+
\bar{\theta}_1^8\bar{\theta}_2^8\right)\right]
\cr
\cr
&&+\left[\theta_3^4-c{\xi_1\choose \xi_2}\left(\theta_4^4+
\theta_1^4\right)-\theta_2^4\right]\left[\bar{\theta}_2^8\bar{\theta}_3^8+
\bar{\theta}_3^8\bar{\theta}_2^8+c{\xi_1\choose \xi_2}
\left(\bar{\theta}_4^8\bar{\theta}_1^8+
\bar{\theta}_1^8\bar{\theta}_4^8\right)\right]
\cr
\cr
&&\left.\left[c{\xi_1\choose \xi_2}\left(\theta_3^4-\theta_1^4\right)-
\theta_4^4-\theta_2^4\right]\left[\bar{\theta}_2^8\bar{\theta}_4^8+
\bar{\theta}_4^8\bar{\theta}_2^8+c{\xi_1\choose \xi_2}
\left(\bar{\theta}_3^8\bar{\theta}_1^8+
\bar{\theta}_1^8\bar{\theta}_3^8\right)\right]\right\}~.\nonumber\\
\ea
In terms of $SO(2n)$ characters $Z_+$ and $Z_-$ are respectively given by
\beq
 Z_+=(V_8-S_8)
\left( \overline O_{16} + \overline S_{16}\right) \left( \overline O_{16} + \overline S_{16}
\right) 
\eeq
and
\ba
 Z_-&=& V_8
\left(\overline O_{16} \overline O_{16}+ \overline C_{16} \overline C_{16}\right) -
S_8 \left( \overline S_{16} \overline S_{16}+
\overline V_{16} \overline V_{16}\right)
\cr
&& + ~ O_8
\left( \overline S_{16} \overline V_{16} + \overline V_{16} \overline S_{16}\right) -C_8
\left( \overline O_{16} \overline C_{16} + \overline C_{16}
\overline O_{16} \right) . 
\ea

The projection implemented in $Z_+$ is 
\beqn
a & = & (-1)^{F+F_{\xi_1}}\,, \nonumber\\
b & = & (-1)^{F+F_{\xi_2}}\,, \label{10dprojection}
\eeqn
with $F$ the space-time fermion number and 
$F_{\xi_{1,2}}$ as before. The choice for the discrete torsion is "$-$". 
The projection in the ten dimensional case reproduces the partition function
of the non--supersymmetric $SO(16)\times SO(16)$ heterotic string.
Thus, we note that in the
free fermion model the same discrete choice of the phase
$c{\xi_1\choose \xi_2}$
is employed in the ten dimensional and four dimensional models. 

\section{Little heterotic strings}\label{shift}
\setcounter{equation}0

In this section we explore other orbifold projections induced by the orbifold
given in eq. (\ref{deltaorbifold}). We comment that the new
partition functions that 
we construct are mathematically consistent in the sense that they
satisfy modular 
invariance and world--sheet supersymmetry. Furthermore, they admit the
Jacobi theta--function identity on the world--sheet supersymmetric side. Therefore,
the spectrum 
possesses $N=4$ space--time supersymmetry and the partition function in
identically zero. 
Consequently, supersymmetric non--renormalization theorems indicate that these
string partition functions are mathematically consistent also at multi--loops.
However, these string solutions are not physical in the sense that the
graviton 
and gauge vector bosons are projected out by the GSO projections. Additionally, 
the conventional spin--statistics assignments, while they hold for massless states, 
are flipped for massive states in the string spectrum. Hence, the partition functions that
we present below do not give rise to physical vacua and corresponding low 
energy effective field theories. Nevertheless, in our judgement they do correspond
to mathematically consistent solutions in the sense articulated above and their
role for a complete understanding of string theory is therefore important.
We comment that in terms of free fermion it is well known that the condition 
$c{{NS} \choose b_j  }=\delta_{b_j}$ guarantees the existence of the graviton in the
spectrum. Analysis of partition functions that do not satisfy this condition
has not been presented to date. 

The partition function of the $E_8\times E_8$ heterotic string in ten dimensions
compactified on a generic six torus is shown in eq. (\ref{e8xe8partin4d}). 
We introduce a shift in one compact dimension $x^9$ 
\begin{equation}
\delta : \quad x^9 \rightarrow x^9 + \pi R, \quad \delta^2=1
\label{x9shift}
\end{equation}
and consider the corresponding shift orbifold generated by the elements
\[(1, (-1)^{F_{\xi_1}}\delta, (-1)^{F_{\xi_2}}\delta, (-1)^{F_{\xi_1}+F_{\xi_2}} )=
(1,a,b,ab)\,,\]
where $F_{\xi_1}$ is the fermion number in the sector describing the first $E_8$
gauge group and  $F_{\xi_2}$
is the fermion number in the sector describing the second  $E_8$ gauge group.
One can consider different options for the signs 
of the elements of the orbifold group.  Different choices will obviously 
lead to different partition functions after orbifolding of the initial model.
We introduce the projection operator
\beq
\frac{1 \mp (-1)^{F_{\xi_1}}\delta}{2}\times \frac{1+(-1)^{F_{\xi_2}}\delta}{2}=
\frac{1}{4}\{1 \mp (-1)^{F_{\xi_1}}\delta+(-1)^{F_{\xi_2}}\delta \mp
(-1)^{F_{\xi_1}+F_{\xi_2}} \}
\label{projections}
\eeq
and proceed with the choice of the minus sign in the first projector of eq. 
(\ref{projections}).
The full partition function which is obtained from (\ref{e8xe8partin4d}) and is 
invariant under the orbifold group is given by
\beq
{Z}_{tot}= {Z}_{oo}+\sum_{i}{Z}_{oi}
+\sum_{i}({Z}_{io}+{Z}_{ii})
+ c_o \sum_{i\ne j} {Z}_{ij}\,,
\label{total}
\eeq
where $i,j\in \{-a,b,-ab\}$ and
the constant $c_o$ is the discrete torsion which can assume values $\pm 1$.
The first two terms in  (\ref{total}) correspond to 
the untwisted sector of the orbifold and are given by
\begin{equation} \label{untw}
{Z}_o=  {Z}_{oo}+\sum_{i}{Z}_{oi}=
{Z}_{o,o}- { Z}_{o,ab} -  {Z}_{o,a} 
+  {Z}_{o,b}
\end{equation}
or explicitly
\ba
{Z}_o=
\frac{1}{4}({V_8} -{S_8})\La_1\La_2\La_{m',n'}\La_{m,n} [ 
&&(\overline O_{16}+\overline S_{16})(\overline O_{16}+\overline S_{16}) \nonumber \\
- && (\overline O_{16}-\overline S_{16})(\overline O_{16}-\overline S_{16})\nonumber\\
+ && (-1)^m\{-(\overline O_{16}-\overline S_{16})
(\overline O_{16}+\overline S_{16})\nonumber\\
+ && (\overline O_{16}+\overline S_{16})( \overline O_{16}-\overline S_{16}) \}]\,.
\label{untwisted}
\ea 
In a similar way the last two terms in (\ref{total}) correspond to the twisted sector
and they read explicitly
\begin{equation} \label{tw1}
\sum_{i}({Z}_{io}+{Z}_{ii})= -{Z}_{a,o}
+ { Z}_{b,o} -  {Z}_{ab,o} 
- {Z}_{a,a}+ { Z}_{b,b} - 
 {Z}_{ab,ab} \,,
\end{equation}

\begin{equation}\label{tw2}
\sum_{i\neq j}{Z}_{i,j}= {Z}_{a,b}+ { Z}_{b,a} + 
{Z}_{ab,a} + {Z}_{ab,b}+ { Z}_{b,ab}
 + {Z}_{a,ab} .
\end{equation}
Below we choose the value of $c_o$ to be $+1$.
Let us note also that if one takes the sign $+$ in the projection (\ref{projections})
then all terms in equations (\ref{untw}), (\ref{tw1}), (\ref{tw2}) which have the sign $-$
will change it to $+$.  
It is straightforward to check that the partition function
(\ref{total}) is invariant under $S$ and $T$ 
transformations, as well under the action of the orbifold group. 
Putting all these into (\ref{total}) we finally obtain
\ba
{Z_-}=({V_8}-{S_8})\La_1\La_2\La_{m',n'}\big[ 
&&\La_{2m,n}~~~~~(\overline S_{16}\overline O_{16}- 
\overline C_{16}\overline C_{16})\nonumber\\
~~~~~~~~~~~~~~~~~~~~~~~~~ 
&+& \La_{2m+1,n}~~~(\overline O_{16}\overline S_{16}- 
\overline V_{16}\overline V_{16})\nonumber\\
~~~~~~~~~~~~~~~~~~~~~~~~~
&+& \La_{2m,n+\frac{1}{2}}~~~(\overline S_{16}\overline C_{16}-
\overline C_{16}\overline O_{16})\nonumber\\
~~~~~~~~~~~~~~~~~~~~~~~~~
&+& \La_{2m+1,n+\frac{1}{2}}(\overline O_{16}\overline V_{16}-
\overline V_{16}\overline S_{16})\big] .
\label{Z-}
\ea
From the expression above one can see that the zero mass spectrum of the model
does not contain the graviton because of the absence of the term $\overline O_{16} \overline O_{16}$
in the right sector. By similar reason there are no gauge degrees of freedom 
either since the adjoint representation of the gauge group has been projected out as well.
Essentially  one ends up with a number of scalar, 
vector and fermion fields in the zero mass sector. 
On the other hand, if one takes the second ($+$) sign in the equation
(\ref{projections}), one obtains the model described by a partition function 
\ba \label{gr}
{Z'_-}=&& ({V_8}-{S_8})\La_1\La_2\La_{m',n'} \big[
~~~
\La_{2m,n}~~~~~(\overline O_{16}\overline O_{16} +\overline C_{16}\overline C_{16})\nonumber\\
&&~~~~~~~~~~~~~~~~~~~~~~~~~
\,\, 
+ \La_{2m+1,n}~~~(\overline S_{16}\overline S_{16}+\overline V_{16}\overline V_{16})\nonumber\\
&&~~~~~~~~~~~~~~~~~~~~~~~~~
\,\, 
+ \La_{2m,n+\frac{1}{2}}~~~(\overline O_{16}\overline C_{16}+
\overline C_{16}\overline O_{16})\nonumber\\
&&~~~~~~~~~~~~~~~~~~~~~~~~~
\,\, 
+ \La_{2m+1,n+\frac{1}{2}}(\overline V_{16}\overline S_{16}+
\overline S_{16}\overline V_{16})\big]~,
\label{ordinaryparti}
\ea
with $c_o=+1$, which contains both gravity and Yang-Mills fields as its massless excitations.
Therefore, we observe that starting from the 
$E_8 \times E_8$ heterotic superstring and performing 
different orbifold projections one can obtain mathematically consistent
heterotic--string solutions with or without gravity and Yang-Mills fields.
We note from eq. (\ref{Z-}) that the second term in each row
has the opposite sign in terms of space--time spin statistics. 
All the states with the "wrong" space--times statistics are
massive string states, whereas the massless states satisfy the 
conventional spin--statistics assignments.
Clearly, therefore, the partition function does not give rise to a
physical vacuum with a local effective field theory limit.
The violation of the spin--statistics assignments indicates the presence of ghosts 
in the spectrum and may signal mathematical inconsistency of this solution for one--loop,
or multi--loop amplitudes. 
However, owing to the Jacobi identity on the supersymmetric side of eq. (\ref{Z-}) 
the spectrum is supersymmetric and tachyon free. Consequently, 
the one--loop partition function in eq. (\ref{Z-})
vanishes exactly, whereas supersymmetric non--renormalization theorems
indicate that multi--loop amplitudes are not renormalized. 
In the next section we 
perform a further $Z_2$ orbifold projection of eq. (\ref{Z-}) 
and discuss the spectra in more details.

As a second example of a little heterotic string we consider the orbifold 
of (\ref{e8xe8partin4d}) by the projection operator given by 
\beq
\frac{1 - (-1)^{F_{\xi_1}}\delta}{2}\times \frac{1-(-1)^{F_{\xi_2}}\delta}{2}=
\frac{1}{4}\{1 - (-1)^{F_{\xi_1}}\delta-(-1)^{F_{\xi_2}}\delta + (-1)^{F_{\xi_1}+F_{\xi_2}} \}\,.
\label{projection2}
\eeq
In this case the resulting partition function for the case with $c_o=-1$ is given by 
\ba
{Z^{''}_-}=&& ({V_8}-{S_8})\La_1\La_2\La_{m',n'} \big[
~~~
\La_{2m+1,n}~~~(\overline O_{16}\overline O_{16} +
                          \overline C_{16}\overline C_{16})\nonumber\\
&&~~~~~~~~~~~~~~~~~~~~~~~~~~
\,\, 
+ ~\La_{2m,n}~~~~~(\overline S_{16}\overline S_{16}+
                            \overline V_{16}\overline V_{16})\nonumber\\
&&~~~~~~~~~~~~~~~~~~~~~~~~~
\,\, 
- ~\La_{2m,n+\frac{1}{2}}~~~(\overline C_{16}\overline O_{16}+ 
                                     \overline O_{16}\overline C_{16})
\nonumber\\
&&~~~~~~~~~~~~~~~~~~~~~~~~~
\,\, 
- ~\La_{2m+1,n+\frac{1}{2}}(\overline S_{16}\overline V_{16}+
                                      \overline V_{16}\overline S_{16})\big].
\label{ordinaryparti2}
\ea
We note from the first line of eq. 
(\ref{ordinaryparti2}) that in this 
case the $\overline O_{16}\overline O_{16}$ term
on the non-supersymmetric side, 
that gives rise to the graviton and gauge vector
bosons, is present. 
However, this term couples to $\La_{2m+1,n}$ in the compactified 
shifted lattice, which is massive at generic points in the moduli space.
and gives rise
to massless states at the enhanced symmetry point with $R=\sqrt{\alpha^\prime}$. 
Also in this case the graviton and the gauge vector bosons 
are not among the massless states.
From the third and fourth rows of eq. (\ref{ordinaryparti2})
we note again that the massive states produced by this partition
function do not satisfy the physical spin--statistics assignments. 
In this model the states with the "wrong" statistics
couple to a lattice with winding modes shifted by $1/2$ and 
hence there are no zero 
modes with the wrong statistics. We note in these examples a link between the 
projection of the graviton and the assignments of "wrong" statistics in the 
massive string spectrum. 
As we argued above, while clearly non--physical, this does 
not signal a mathematical inconsistency of these solutions for 
one-- or multi--loop 
amplitudes, due to existence of supersymmetry in the spectrum, 
owing to the Jacobi
identity on the supersymmetric side. One can further contemplate in the case of 
(\ref{ordinaryparti2}) the breaking of $N=4$ to $N=2$ and $N=1$ by 
$Z_2$ twists of the internal lattice and the possibility that the resulting
spectrum in the twisted sectors is physically sensible, {\it i.e.} that the
massless twisted states satisfy the conventional spin--statistics assignments.

\section{$Z_2$ projection} 
\setcounter{equation}0

In this section we consider the $Z_2$ orbifold of the partition 
function (\ref{Z-}). The $Z_2$ is generated by the elements $(1,h)$
where $h$ acts on the coordinates of the internal tori
as 
\[Z_1\rightarrow e^{i\pi}Z_1\,\,\,\,\,\,\,\,\,,\,\,\,\,\
Z_2\rightarrow e^{i\pi}Z_2\,\,\,\,\,\,,\,\,\,\,\,
Z_3\rightarrow Z_3 \,\,\,.
\]
As in the case of the standard embedding the element
$h$ acts non-trivially on the gauge degrees of freedom of the heterotic string
as well. 
For this reason it is convenient to decompose the $SO(2n)$ characters
in such a way to keep $O_4,V_4,S_4 $ and $C_4$ factors 
(on which the element $h$ acts
non-trivially) explicit.
In particular the $Z_{oo}$ part of the partition function or equivalently
the partition function (\ref{Z-}) can be rewritten in the form
\ba \label{oo}
{Z_{\text{oo}}}&=& \frac{1}{4}\big[
 V_4  O_4+ O_4  V_4- S_4 S_4- C_4 
 C_4\big]
\La_1\La_2\La_{m',n'} \\ 
&& \times \{(\La_{2m,n}+\La_{2m,n+\frac{1}{2}})
(\overline S_4\overline S_{12}+
\overline C_4\overline C_{12}-\overline S_4 \overline C_{12}-
\overline C_4\overline S_{12})(\overline O_{16}
+\overline C_{16})  \nonumber \\
&&+(\La_{2m,n}-\La_{2m,n+\frac{1}{2}})
(\overline S_4\overline S_{12}+\overline C_4\overline C_{12}+
\overline S_4\overline C_{12}+
\overline C_4\overline S_{12})(\overline O_{16}-\overline C_{16}) \nonumber \\
&&+(\La_{2m+1,n}+\La_{2m+1,n+\frac{1}{2}})
(\overline O_4\overline O_{12}+
\overline V_4\overline V_{12}-\overline V_4\overline O_{12}-
\overline O_4\overline V_{12})
(\overline S_{16}+\overline V_{16}) \nonumber \\
&&+ (\La_{2m+1,n}-
\La_{2m+1,n+\frac{1}{2}})
(\overline O_4\overline O_{12}+\overline V_4\overline V_{12}
+\overline V_4\overline O_{12}+\overline O_4\overline V_{12})
(\overline S_{16}-\overline V_{16})
\}.\nonumber 
\ea
In order to obtain $Z_{\text{oh}}= hZ_{\text{oo}}$ let us recall how the operator $h$ acts
on the $SO(2n)$ characters. In particular we have
\beq
O_4\rightarrow O_4 ~~ ,\hspace{1cm} V_4\rightarrow -V_4 ~~ ,\hspace{1cm}
S_4\rightarrow -S_4 ~~ ,\hspace{1cm} C_4\rightarrow C_4\nonumber
\eeq
and 
\ba
O_{16}&=& O_4O_{12}+V_4V_{12} \rightarrow +O_4O_{12}-V_4V_{12} ,\nonumber\\
V_{16}&=& V_4O_{12}+O_4V_{12} \rightarrow -V_4O_{12}+O_4V_{12} ,\nonumber\\
S_{16}&=& S_4S_{12}+C_4C_{12} \rightarrow -S_4S_{12}+C_4C_{12} ,\nonumber\\
C_{16}&=& S_4C_{12}+C_4S_{12} \rightarrow -S_4C_{12}+C_4S_{12} .\nonumber
\ea
We also need to take into account that  $h (\La_{1}\La_2)\rightarrow |\frac{2 \eta}
{\theta_2}|^4$.
Therefore we obtain
\ba \label{oh}
{Z_{\text{oh}}}&=& \frac{1}{4}\big[
- V_4 O_4+ O_4  V_4+S_4  S_4- C_4  C_4\big]
\La_{m',n'}|\frac{2 \eta}
{\theta_2}|^4  \\
&&\times \big\{(\La_{2m,n}+\La_{2m,n+\frac{1}{2}})
(-\overline S_4\overline S_{12}+
\overline C_4\overline C_{12}+\overline S_4\overline C_{12}-
\overline C_4\overline S_{12})(\overline O_{16}+\overline C_{16})\nonumber\\
&&+(\La_{2m,n}-\La_{2m,n+\frac{1}{2}})
(-\overline S_4\overline S_{12}+\overline C_4\overline C_{12}-
\overline S_4\overline C_{12}+\overline C_4\overline S_{12})
(\overline O_{16}-\overline C_{16})\nonumber\\
&&+(\La_{2m+1,n}+\La_{2m+1,n+\frac{1}{2}})
(\overline O_4\overline O_{12}-\overline V_4\overline V_{12}
+\overline V_4\overline O_{12}-\overline O_4\overline V_{12})
(\overline V_{16}+\overline S_{16}) \nonumber \\
&&+ (\La_{2m+1,n}-\La_{2m+1,n+\frac{1}{2}})
(\overline O_4\overline O_{12}-\overline V_4\overline V_{12}
-\overline V_4\overline O_{12}+\overline O_4\overline V_{12})
(\overline S_{16}-\overline V_{16})
\big\} . \nonumber
\ea
The massless contributions are given by
\beq
 \frac{\overline C_4\overline C_{12}\overline O_{16}}
{\overline\eta^8} \times \frac{ O_4  V_4 - C_4  C_4 }{ \eta^8}, \quad
 \frac{\overline S_4\overline S_{12}\overline O_{16}}{\overline\eta^8}
 \times \frac{ V_4  O_4 - S_4  S_4 }{ \eta^8},
\eeq
since left and right contributions give
$$
 \frac{\overline C_4\overline C_{12}\overline O_{16}}{\overline\eta^8}
\sim 2^6\overline q^0  ,\quad  \frac{\overline S_4\overline S_{12}\overline O_{16}}
{\overline\eta^8} \sim 2^6 \overline q^0 , 
$$

$$
\frac{O_4 V_4  }{ \eta^8} \sim 4 q^0, \quad
\frac{ V_4  O_4  }{ \eta^8} \sim 4  q^0, \quad
\frac{S_4  S_4  }{ \eta^8} \sim 4 q^0, \quad
\frac{ C_4  C_4  }{ \eta^8} \sim 4  q^0. 
$$
From the expressions above one can read the content of
the massless spectrum in terms of the 
six dimensional $N = (1,1)$ (which gives $D=4$ $N=4$ supersymmetry upon 
the dimensional reduction to four dimensions) SUSY multiplets.
In particular one has $2^6$ massless $(1,1)$ multiplets, whose bosonic
part contains one vector and four scalar fields.

In order to obtain the partition function in the  twisted sector
one should apply the chain of transformations
\[{Z_{\text{ho}}}= S {Z_{\text{oh}}}\,\,\,\,\,\,,\,\,\,\,\,\,
{Z_{\text{hh}}}= T {Z_{\text{ho}}}\,.
\]
Doing so, one finally obtains\footnote{In order to perform $S$ transformations it 
is convenient to rewrite $CC-SS=\frac{1}{2}((C-S)(C+S)+ (C+S)(C-S))$ etc.}
\ba \label{ho}
{Z_{\text{ho}}}&=& \frac{1}{4}\big[
- V_4  C_4+ S_4  V_4- O_4  S_4+ C_4 O_4\big]
\La_{m',n'}\times 16|\frac{\eta}{\theta_4}|^4  \\
&& \times \big\{(\La_{2m,n}
+\La_{2m,n+\frac{1}{2}})
(\overline O_4\overline S_{12}-
\overline O_4\overline C_{12}-\overline V_4\overline S_{12}+
\overline V_4\overline C_{12})(\overline O_{16}+\overline C_{16}) \nonumber \\
&&+(\La_{2m+1,n}+\La_{2m+1,n+\frac{1}{2}})
(\overline S_4\overline O_{12}-\overline S_4\overline V_{12}+
\overline C_4\overline V_{12}-\overline C_4\overline O_{12})
(\overline S_{16}+\overline V_{16}) \nonumber \\
&&+(\La_{2m,n}-\La_{2m,n+\frac{1}{2}})
(\overline O_4\overline S_{12}+\overline O_4\overline C_{12}+
\overline V_4\overline S_{12}+\overline V_4\overline C_{12})
(\overline O_{16}-\overline C_{16}) \nonumber \\
&&+(\La_{2m+1,n}-\La_{2m+1,n+\frac{1}{2}})
(\overline S_4\overline O_{12}+\overline S_4\overline V_{12}+
\overline C_4\overline O_{12}+\overline C_4\overline V_{12})
(\overline S_{16}-\overline V_{16})
\big\} ,\nonumber
\ea
\ba \label{hh}
{Z_{\text{hh}}}&=& \frac{1}{4}\big[
- V_4  C_4+ S_4  V_4+O_4  S_4- C_4 O_4\big]
\La_{m',n'}\times 16|\frac{\eta}{\theta_3}|^4 \\
&&\times \big\{(\La_{2m,n}+\La_{2m,n+\frac{1}{2}})
(-\overline O_4\overline S_{12}+\overline O_4\overline C_{12}
-\overline V_4\overline S_{12}+\overline V_4\overline C_{12})
(\overline O_{16}+\overline C_{16}) \nonumber \\ 
&&+(\La_{2m+1,n}-\La_{2m+1,n+\frac{1}{2}})
(\overline S_4\overline O_{12}+\overline S_4\overline V_{12}
-\overline C_4\overline V_{12}-\overline C_4\overline O_{12})
(\overline S_{16}-\overline V_{16}) \nonumber \\ 
&&+(\La_{2m,n}-\La_{2m,n+\frac{1}{2}})
(-\overline O_4\overline S_{12}-\overline O_4\overline C_{12}
+\overline V_4\overline S_{12}+\overline V_4\overline C_{12})
(\overline O_{16}-\overline C_{16}) \nonumber \\  
&&+(\La_{2m+1,n}+\La_{2m+1,n+\frac{1}{2}})
(\overline S_4\overline O_{12}-\overline S_4\overline V_{12}
+\overline C_4\overline O_{12}-\overline C_4\overline V_{12})
(\overline S_{16}+\overline V_{16})
\big\} .\nonumber
\ea
The massless contributions are given by 
\beq
 \frac{\overline O_4\overline S_{12}\overline O_{16}}
{\overline\eta^2\bar \theta_4^2} \times \frac{O_4  S_4 - C_4  O_4 }
{\eta^2\theta^2_4} ,
 \eeq
since left and right contributions give

$$
 \frac{\overline C_4\overline C_{12}\overline O_{16}}
{\overline\eta^2\bar\theta_4^2}\sim 2^5 \overline q^0, \quad  
\frac{ O_4  S_4  }{\eta^2\theta^2_4} \sim 2  q^0, \quad
\frac{ C_4  O_4  }{ \eta^2\theta^2_4} \sim 2  q^0. 
$$
As it was for the case of the untwisted sector one can group the massless spectrum in terms on six dimensional supersymmetry
multiplets. In the twisted sector  one has $D=6$ $N=1$ supersymmetry (which gives $D=4$ $N=2$ upon
the reduction to four dimensions)  and the massless spectrum  forms  $2^5$ half-hypermultiplets.

Repeating similar steps for the model (\ref{gr})
one obtains the following massless contributions. In the untwisted sector
 one has $N=2$ $D=6$ SUGRA multiplet and Yang--Mills multiplet
\beq
 \frac{\overline O_4\overline O_{12}\overline O_{16}}
{\overline\eta^8} \times \frac{ O_4  V_4 - C_4  C_4 }{ \eta^8}, \quad
 \frac{\overline V_4\overline V_{12}\overline O_{16}}{\overline\eta^8}
 \times \frac{ V_4  O_4 - S_4  S_4 }{ \eta^8},
\eeq
\beq
 \frac{\overline O_4\overline V_{12}\overline V_{16}}
{\overline\eta^8} \times \frac{ O_4  V_4 - C_4  C_4 }{ \eta^8}, \quad
 \frac{\overline V_4\overline O_{12}\overline V_{16}}{\overline\eta^8}
 \times \frac{ V_4  O_4 - S_4  S_4 }{ \eta^8}
\eeq
and in the twisted sector one has $N=1$, $D=6$ half hypermultiplets
\beq
 \frac{\overline C_4\overline V_{12}\overline O_{16}}
{\overline\eta^2\bar\theta^2_4} \times \frac{ O_4  S_4 - C_4  O_4 }{ \eta^2\theta_4^2}, \quad
 \frac{\overline S_4\overline O_{12}\overline O_{16}}{\overline\eta^2\bar\theta^2_4}
 \times \frac{ O_4  S_4 - C_4  O_4 }{ \eta^2\theta^2_4}.
\eeq

\section{String Thermodynamics}
\setcounter{equation}0
In order to study further various properties of the 
model such as the asymptotic 
density of the states, 
the Hagedorn transitions etc., one can consider the string thermodynamics
at nonzero temperature 
(see for example \cite{McClain:1986id,O'Brien:1987pn,Polchinski:1985zf,Atick:1988si,Bowick:1989us,Deo:1988jj} 
and references therein). 
This is usually done by considering a propagation of the superstring
with the time coordinate being Wick rotated and compactified on a circle. 
The radius $\beta$ of the compactified Euclidean time is inverse to the temperature
$T$, i.e., 
\begin{equation}
X^0 (\sigma, \tau)= x^0 + \frac{2 \pi m \tau}{\beta} + \frac{n \beta \sigma}{\pi} + osc.
\end{equation}
The corresponding free energy  is expressed as an integral over the fundamental
domain of the torus
 \cite{Atick:1988si}
\begin{equation}
\frac{F}{V} =  \frac{1}{4} {(\frac{1}{ 4 \pi \alpha^\prime})}^5 
\int \frac{d^2 \tau}{ \tau_2^6} \sum_{n,m \in Z} 
({{\cal V}_8 (n,m)} 
-{{\cal S}_8(n,m)})(\overline O_{16}+\overline S_{16})
(\overline O_{16}+\overline S_{16}) e^{- S_\beta (n,m)} .
\label{Z+0T}
\end{equation}
Here $V$ is a spatial volume and the usual $SO(8)$ characters are modified according to
\begin{equation}
{{\cal V}_8 (n,m)}   =\frac{\te^4_3 U_3(n,m) -\te^4_4 U_4(n,m)}{2\eta^4}, \quad
{{\cal S}_8(n,m)}   =\frac{\te^4_2 U_2(n,m) +\te^4_1 U_1(n,m)}{2\eta^4},
\end{equation}
where
\begin{equation}
U_1(n,m)= \frac{1}{2}(-1 + {(-1)}^n + {(-1)}^m + {(-1)}^{n+m}  ),
\end{equation} 
$$
U_2(n,m)= \frac{1}{2}(1 - {(-1)}^n + {(-1)}^m + {(-1)}^{n+m}  ),
$$ 
$$
U_3(n,m)= \frac{1}{2}(1 + {(-1)}^n + {(-1)}^m - {(-1)}^{n+m}  ),
$$ 
$$
U_4(n,m)= \frac{1}{2}(1 + {(-1)}^n - {(-1)}^m + {(-1)}^{n+m}  ),
$$ 
and
\begin{equation}
S_\beta(n,m)= \frac{\beta^2}{4 \pi \alpha^\prime \tau_2}(m^2 + n^2 {|\tau|}^2 - 2 \tau_1 nm) .
\end{equation}

The modular invariance of  
(\ref{Z+0T})
can be easily checked using the rules
\ba
&&U_1(n,m)~~ \underrightarrow{S} ~~U_1(m,-n)~~~~~ ,~~  
U_2(n,m)~~\underrightarrow{S} ~~U_4(m,-n)~~~~~,\nonumber\\
&& U_3(n,m)~~\underrightarrow{S}~~ U_3(m,-n)~~~~~,
~~U_4(n,m)~~\underrightarrow{S} ~~U_2(m,-n)~~~~~ ,  \nonumber\\
&&U_1(n,m)~~\underrightarrow{T}~~ U_1(n,m-n)~~ ,~~ 
U_2(n,m)~~\underrightarrow{T}~~ U_2(n,m-n) ~~,\nonumber\\
&&  U_3(n,m)~~\underrightarrow{T}~~ U_4(n,m-n)~~ ,~~ 
U_4(n,m)~~\underrightarrow{T}~~ U_3(n,m-n)~~,  \nonumber\\
&&  S_\beta (n,m) ~~\underrightarrow{S}~~ S_\beta(m,-n)~~~~~ ,
~~S_\beta (n,m)~~ \underrightarrow{T}~~ S_\beta(n,m-n)~~.\nonumber\\
\ea
As a result of the compactification on  the circle the mass-shell
condition for the heterotic string
 gets modified by the terms which involve the temperature
\begin{equation}
\frac{M^2_R}{8}= N - \frac{1}{2} +\frac{1}{2}{(\frac{\pi \tilde m}{\beta}- 
\frac{n \beta}{2 \pi})}^2, \quad
 \frac{M_L^2}{8}= \tilde N - 1 +\frac{1}{2}{(\frac{\pi \tilde m}{\beta}+ 
\frac{n \beta}{2 \pi})}^2 ,
\end{equation}
where $\tilde m$ is allowed to assume also half-integer values  \cite{Atick:1988si}.
As a result of the presence of the functions $U_i(n,m)$ in the 
partition functions, the usual GSO projection gets modified
and certain states which have been projected out at zero 
temperature may reappear, such as for example $N= \tilde N=0$
and $n=1, \tilde m = \frac{1}{2}$.
Due to the dependence on  $\beta$ in the mass-shell equations  
these states become tachyonic at certain temperature (Hagedorn transition)
indicating the instability of the system at high energies.

Now one can follow the same lines as for the case  of  $T=0$. 
Let us note that the  modification
comparing to the case of $T=0$ is  
$(V_8 - S_8) \rightarrow 
({\cal V}_8 - {\cal S}_8)e^{-S_\beta}$, whereas
the right part of the partition function with nonzero temperature is the same as
in (\ref{oh}), (\ref{ho}), (\ref{hh}) 
and we shall not spell it out explicitly.
After  applying consequently  $Z_2$, $S$ and $T$ transformations on 
$({\cal V}_8 - {\cal S}_8)$ one gets therefore:

$\bullet$ For ${Z_{\text{oo}}}$ in eq. (\ref{oo}),  
$\big[
 V_4  O_4+ O_4  V_4- S_4  S_4- C_4 
C_4\big]$ is substituted by
\begin{equation}
 \frac{1}{2 \eta^4} \sum_{m,n}(U_3 \te_3^2 \te_3^2 - U_4 \te_4^2 \te_4^2 -
U_2 \te_2^2 \te_2^2 - U_1 \te_1^2 \te_1^2) e^{-S_\beta}
\end{equation}

$\bullet$ For ${Z_{\text{oh}}}$ in eq. (\ref{oh}), 
$\big[-
 V_4  O_4+O_4  V_4+ S_4  S_4- C_4 
 C_4\big]$ is substituted by
\begin{equation}
\frac{1}{2 \eta^4} \sum_{m,n}(U_3 \te_3^2 \te_3^2 - U_4 \te_4^2 \te_4^2 -
U_2 \te_2^2 \te_2^2 - U_1 \te_1^2 \te_1^2) e^{-S_\beta}
\end{equation}

$\bullet$ For ${Z_{\text{ho}}}$ in eq. (\ref{ho}),  
$\big[-
 V_4  C_4+ S_4  V_4-O_4  S_4+ C_4 
O_4\big]$ is substituted by
\begin{equation}
 \frac{1}{2 \eta^4} \sum_{m,n}(U_3 \te_2^2 \te_3^2 - U_2 \te_3^2 \te_2^2 -
U_4 \te_1^2 \te_4^2 - U_1 \te_4^2 \te_1^2) e^{-S_\beta}
\end{equation}

$\bullet$ For ${Z_{\text{hh}}}$ in eq. (\ref{hh}), 
$\big[-
V_4  C_4+S_4  V_4+ O_4 S_4- C_4 
 C_4\big]$ is substituted by
\begin{equation}
 \frac{1}{2 \eta^4} \sum_{m,n}(U_3 \te_2^2 \te_4^2 - 
U_4 \te_4^2 \te_2^2 -U_2 \te_1^2 \te_3^2 - U_1 \te_3^2 \te_1^2) e^{-S_\beta}.
\end{equation}
Now having the partition function  one can 
study in detail the corresponding 
high energy thermodynamics following the lines of 
\cite{Deo:1988jj}. On the other hand,
having introduced the temperature dependence into the
partition function, one can consider a more general orbifold projection 
compared to the one which has been done in the section \ref{shift}.
Namely, one can make  the sign in the projection 
(\ref{projections}) to depend on the integers $n$ and $m$,
which corresponds to the "compactified" time direction.
In other words, one can introduce the temperature dependence
into the orbifold projection as well.
In this case one can consider a certain duality between
two a priory different models considered above.
A particular choice of the $+$ sign 
in (\ref{projections})
if both $m$ and $n$ are even and of the $-$ sign if at least one of $m$ and $n$ is
odd leads  to the modular invariant partition function which is a
superposition of the partition functions
${Z'_-}$ (for $m,n$ even) and ${Z_-}$ (for  at least one of $m,n$ is odd).
In this particular model one has the  gravity and Yang-Mills fields from 
the model described by
${Z'_-}$, whereas the massive excitations are described by the ones of both
${Z'_-}$ and ${Z_-}$.

\section{Conclusions}\label{conclude}

In this paper we studied orbifolds of the heterotic $E_8\times E_8$ string
in which the graviton is projected out.
Our starting point is the partition function of the  $E_8 \times E_8$
heterotic--string compactified on a six torus $T^2 \times T^2 \times T^2$.
Our $Z_2\times Z_2$ orbifold includes a specific shift-- orbifold projection
which acts on an internal circle coupled with the fermion numbers of the 
gauge degrees of freedom. The action of this orbifold, in general, 
results in the breaking of $E_8\times E_8$ to $SO(16)\times SO(16)$ and reproduces
the equivalent breaking in the quasi--realistic free fermionic models, where it is 
realised by a discrete choice of a generalised GSO phase.  
Our paper clarifies in this sense the relation between free fermion models
and orbifold models. The breaking induced by the orbifold of eq. (\ref{deltaorbifold})
can be implemented in constructions of quasi--realistic orbifold models.
The phenomenological advantages that it may offer over conventional Wilson--line breaking
is in the Higgs--matter splitting \cite{higgsmattersplit}.
We then studied variations of this $Z_2\times Z_2$ orbifold by modifying the 
sign of the orbifold and constructing the modular invariant partition function.
Additionally, we studied a $Z_2$ projection which acts on four out of six 
coordinates of the compactified torus. 

Depending on the sign of the generator of the orbifold groups one gets different models.
The resulting models are modular invariant. They possess $N=4$, or $N=2$, space--time
supersymmetry and  have rather different low energy spectra. While the 
first contains the graviton and the gauge vector bosons
among its massless excitation, the second model has neither graviton nor Yang-Mills fields
in the massless spectrum, anywhere in the moduli space, owning to the absence of the 
$\overline O_{16}\overline O_{16}$ character in the partition function. In the third model that
we introduced the $\overline O_{16}\overline O_{16}$ character does appear in the partition, but couples to 
the lattice term $\La_{2m+1,n}$, which is massive at a generic point in the moduli 
space. It is tempting to conjecture a closer connection of the latter two models
with the little string theory, where the graviton is also consistently
decoupled, without taking the low energy limit $\alpha^\prime \rightarrow 0$.
A more detailed study of the thermodynamics of the model could hopefully give
some more insight into a possible connection with little string theories, where 
the high energy thermodynamics has been studied extensively
\cite{Kutasov:2000jp,Harmark:2000hw}.
We have further seen that the massive spectrum in the 
resulting partition functions does not admit the conventional spin--statistics 
assignments.
In the third model this is clearly the case, as the terms with the "wrong" statistics
couple to a lattice term, with winding modes shifted by $1/2$, whereas in the second
case closer examination of the relevant characters shows that this is indeed the case.
We therefore see a connection between the projection of the
graviton and the reversal of
the usual spin--statistics assignments in the massive string spectrum.
Curiously enough, 
the massless terms still preserve the conventional spin--statistics
assignments. 
Clearly, the absence of the graviton and the gauge vector bosons, as well as 
the mismatch of the spin--statistics, mean that these models are not physical 
vacua. We articulated that this indicates the existence of
ghosts in these models, but owning to
the existence of the Jacobi identity on the supersymmetric side and the 
supersymmetric 
nonrenormalisation theorems, their partition functions
correspond to mathematically 
consistent solutions for one--, as well as multi--loop, amplitudes. 
One may question whether
the correct spin--statistics assignments
need to hold for the massive string spectrum.
The change of the conventional spin--statistics 
assignments in the massive string spectrum may indicate
the breaking of Lorentz covariance, a tenant
of local quantum field theories, above the string scale.

We briefly discussed the generalization of our orbifold models to the case when
the orbifold projection depends on the temperature. This generalization
allows in a certain sense 
to combine the two models under the consideration. In this case the gravity
and Yang-Mills theories
appear at the low energy spectrum while the massive excitations are
described by the ones of both models.

The string solutions presented in this paper are clearly
non--conventional. In string
constructions to date the correct spin--statistics 
assignment for all string states is
taken as a pivotal requirement. However, our view is that 
our understanding of string
theory is still very rudimentary. Its varying features should be explored and 
``conventional wisdom'' be challenged in order to gain deeper insight.
The existence of mathematically consistent string solutions, 
which do not contain gravity,
possibly coupled with a breakdown of Lorentz covariance above the 
string scale, suggests an
interesting scenario. Namely, gravity and local gauge theories as
emerging effective field theories below the string scale. 

\bigskip
\medskip
\leftline{\large\bf Acknowledgements}
\medskip

We would like to thank Carlo Angelantonj, Thomas Mohaupt, Mirian Tsulaia and
Cristina Timirgaziu for useful discussions. 
This work was supported by the STFC  and the University of Liverpool.




\bibliographystyle{unsrt}

\begin{thebibliography}{99}

\bibitem{Hull:1994ys}
  C.M. Hull and P.K.~Townsend, \NPB{438}{1995}{109}.

\bibitem{Witten:1995ex}
  E.~Witten, \NPB{443}{1995}{85}.

\bibitem{Dixon:1986iz}
  L.~J.~Dixon and J.~A.~Harvey, \NPB{274}{1986}{93}.

\bibitem{lewellen}
              H. Kawai, D.C. Lewellen, and S.H.-H. Tye, \PRD{34}{1986}{3794}.

\bibitem{nillesraby} O. Lebedev \etal, \PRD{77}{2008}{046013}. 

\bibitem{higgsmattersplit} \AEF, \EJP{49}{2007}{803}.

\bibitem{partitions}
  A.E.~Faraggi, \PLB{544}{2002}{207}.


\bibitem{Kutasov:2000jp}
  D.~Kutasov and D.~A.~Sahakyan,  JHEP {\bf 0102} (2001) 021.

\bibitem{Harmark:2000hw}
  T.~Harmark and N.A.~Obers, \PLB{485}{2000}{285}.

\bibitem{Aharony:1999ks}
  O.~Aharony,
  Class.\ Quant.\ Grav.\  {\bf 17} (2000) 929.

\bibitem{Kutasov:2001uf}
  D.~Kutasov, {\it Prepared for ICTP Spring School on Superstrings 
and Related Matters, Trieste, Italy, 2-10 Apr 2001}


\bibitem{spduality} \AEF, C. Kounnas and J. Rizos, \NPB{714}{2007}{208}; 
					           \NPB{799}{2008}{19}; \\
T. Catelin-Jullien, A.E. Faraggi, C. Kounnas and J. Rizos, 
							\NPB{812}{2009}{103}.

\bibitem{fff} I. Antoniadis, C. Bachas, and C. Kounnas, \NPB{289}{1987}{87};\\
               H. Kawai, D.C. Lewellen, and S.H.-H. Tye, \NPB{288}{1987}{1}.

\bibitem{nahe} \AEF~and D.V. Nanopoulos, \PRD{48}{1993}{3288}.

\bibitem{ffmreview} For a recent review see {\it e.g.}: \\
\AEF,  arXiv:0809.2641, and references therein. 

\bibitem{foc} \AEF, \PLB{326}{1994}{62}.

\bibitem{Narain} K. Narain, \PLB{169}{1986}{41};\\
                 K.S. Narain, M.H. Sarmadi and E. Witten,
						\NPB{279}{1987}{369}.

\bibitem{Angelantonj:2002ct}
  C.~Angelantonj and A.~Sagnotti,
  Phys.\ Rept.\  {\bf 371} (2002) 1.

\bibitem{McClain:1986id}
  B.~McClain and B.~D.~B.~Roth,
  Commun.\ Math.\ Phys.\  {\bf 111} (1987) 539.

\bibitem{O'Brien:1987pn}
  K.~H.~O'Brien and C.~I.~Tan, \PRD{36}{1987}{1184}.

\bibitem{Polchinski:1985zf}
  J.~Polchinski,
  Commun.\ Math.\ Phys.\  {\bf 104} (1986) 37.

\bibitem{Atick:1988si}
  J.~J.~Atick and E.~Witten, \NPB{310}{1988}{291}.

\bibitem{Bowick:1989us}
  M.~J.~Bowick and S.~B.~Giddings, \NPB{325}{1989}{631}.

\bibitem{Deo:1988jj}
  N.~Deo, S.~Jain and C.~I.~Tan, \PLB{220}{1989}{125}.










\end{thebibliography}

\vfill\eject
\end{document}